

\newif\iffigs\figsfalse
\figstrue

\input harvmac
\let\tilde\widetilde

\iffigs
  \input epsf
\else
  \message{No figures will be included. See TeX file for more
information.}
\fi
\catcode`\@=11
\newbox\Bbox
\newif\ifBinsert
\newdimen\fullvsize \fullvsize=\vsize
\newtoks\dfoutput \dfoutput=\the\output
\ifx\answ\bigans\else
\output={
  \ifBinsert
    \if L\l@r
       \the\dfoutput
    \else
       \count1=2
       \message{Doing wide insert.}
       \shipout\vbox{\speclscape
          {\hsize=\fullhsize \makeheadline}
          \hbox to\fullhsize{\hfil\box\Bbox\hfil}
          \hbox to\fullhsize{\box\leftpage\hfil
                \leftline{\vbox{\pagebody\makefootline}}}}
       \global\let\l@r=L
       \advancepageno
       \global\Binsertfalse
       \global\vsize=\fullvsize
    \fi
  \else
    \if B\botmark
      \if R\l@r
         \mark{}
         \global\Binserttrue
       \the\dfoutput
         \global\advance\vsize by -\ht\Bbox
      \else
         \the\dfoutput
      \fi
    \else
      \the\dfoutput
    \fi
  \fi
}
\fi
\def\Title#1#2{\nopagenumbers\abstractfont\hsize=\hstitle\rightline{#1}%
\vskip 0.5in\centerline{\titlefont #2}\abstractfont\vskip .5in\pageno=0}

\font\eightrm=cmr8 \font\eighti=cmmi8
\font\eightsy=cmsy8 \font\eightbf=cmbx8
\font\eightit=cmti8 \font\eightsl=cmsl8 \skewchar\eighti='177
\skewchar\eightsy='60

\def\eightpoint{\def\rm{\fam0\eightrm}
\textfont0=\eightrm \scriptfont0=\fiverm \scriptscriptfont0=\fiverm
\textfont1=\eighti \scriptfont1=\fivei \scriptscriptfont1=\fivei
\textfont2=\eightsy \scriptfont2=\fivesy \scriptscriptfont2=\fivesy
\textfont\itfam=\eighti
\def\it{\fam\itfam\eightit}\def\sl{\fam\slfam\eightsl}%
\textfont\bffam=\eightbf \def\bf{\fam\bffam\eightbf}\rm}

\def\inbar{\,\vrule height1.5ex width.4pt depth0pt}
\font\cmss=cmss10 \font\cmsss=cmss8 at 8pt
\def\BZ{\relax\ifmmode\mathchoice
{\hbox{\cmss Z\kern-.4em Z}}{\hbox{\cmss Z\kern-.4em Z}}
{\lower.9pt\hbox{\cmsss Z\kern-.36em Z}}
{\lower1.2pt\hbox{\cmsss Z\kern-.36em Z}}\else{\cmss Z\kern-.4em Z}\fi}
\def\IC{\relax\hbox{$\inbar\kern-.3em{\rm C}$}}
\def\IP{\relax{\rm I\kern-.18em P}}
\def\IQ{\relax\hbox{$\inbar\kern-.3em{\rm Q}$}}
\def\IR{\relax{\rm I\kern-.18em R}}
\let\IZ\BZ
\def\CFT{conformal field theory}
\def\CY{Calabi-Yau}
\def\K{K\"ahler}
\def\NL{nonlinear sigma model}

\def\ex#1{\hbox{$\> e^{#1}\>$}}
\def\CP#1{{\IP}^{#1}}

\Title{\vbox{\baselineskip12pt\hbox{IASSNS-HEP-93/81}}}
{\vbox{\centerline{Spacetime Topology Change:}
    \vskip2pt\centerline{The Physics of Calabi-Yau Moduli Space$^*$}}}
\vfootnote{$^*$}{Lecture delivered by B.R.G. at Strings '93, Berkeley.}

\centerline{Paul S. Aspinwall,%
\footnote{${}^\dagger$}{
School of Natural Sciences, Institute for Advanced Study,
Princeton, NJ \ 08540.
}
Brian R. Greene%
\footnote{${}^\ddagger$}{
School of Natural Sciences, Institute for Advanced Study,
Princeton, NJ \ 08540.
On leave from:
F.R. Newman Laboratory of Nuclear Studies,
Cornell University, Ithaca, NY \ 14853.
}
and David R. Morrison%
\footnote{${}^{\mathchar "278}$}{
School of Mathematics, Institute for Advanced Study,
Princeton, NJ \ 08540.
On leave from:  Department of Mathematics, Duke University,
Box 90320, Durham, NC \ 27708.}}

\vskip 1 in

\noblackbox

We review recent work which has significantly sharpened our
geometric understanding and interpretation of the moduli space
of certain $N$=2 superconformal field theories.  This has
resolved some important issues in mirror symmetry and
has also established that string theory admits physically smooth
processes which can result in a change in topology of the
spatial universe.

\Date{11/93}

\newsec{Introduction}
The essential lesson of general relativity is that the geometrical
structure of spacetime is governed by dynamical variables. That is,
the metric  changes in time according to the Einstein
equations. In the usual formulations of general relativity, the spacetime
metric is defined on a space of fixed topological type -- the ``size'' and
``shape'' of the space can smoothly change, but the underlying topology
does not. A natural question to ask is whether this formulation is
too restrictive; might the topology of space
itself be a dynamical variable and hence possibly change in time?
This issue has long been speculated upon. Heuristically,
one suspects that topology might be able to change by means of the violent
curvature fluctuations which would be expected in any quantum theory of
gravity.
Just as the fluctuations of the magnetic field in a box of size
$L$ are on the order of $(\hbar c)^{1/2}/L^2$,
 those of the curvature of the gravitational field are on the
order of $({\hbar G \over c^3})^{1/2}/L^3$.
 Thus, on extremely small scales, say $L \sim
L_{\rm Planck}$, huge curvature fluctuations are unsuppressed.
One can imagine that such curvature fluctuations could ``tear'' the fabric of
space resulting in a change of topology. The expected discontinuities
in physical observables accompanying the discontinuous operation
of a change in topology would be hidden, one hopes, behind the smoothing
effects
of quantum uncertainty. Of course, without a true theory of quantum
gravity, one cannot make quantitative sense of such hypothesized
processes.

\nref\rW{E. Witten, Nucl. Phys. {\bf B403} (1993) 159.}
\nref\rAGM{P.S. Aspinwall, B.R. Greene and D.R. Morrison,
Phys. Lett.  {\bf 303B} (1993) 249\semi
``Calabi-Yau moduli space, mirror manifolds and spacetime
topology change in string theory'',
IASSNS-HEP-93/38, CLNS-93/1236.}
With the advent of string theory, we are led to ask whether any new
quantitative light is shed on the issue of topology change. Two works
over the last year \refs{\rW,\rAGM}
have carried out studies, from somewhat different points
of view, which definitively establish that there are {\it physically smooth}\/
processes in string theory which result in a change in the topology
of spacetime.\foot{Recently, another example illustrating
topology change in string theory has  been  proposed
\ref\rKG{A.Giveon and E.Kiritsis, ``Axial Vector Duality as a Gauge
Symmetry and Topology Change in String Theory'', hep-th/9303016.}.}
Furthermore, as phenomena in string
theory, these processes are not at all exotic.
Rather, they correspond to the most basic kind of operation
arising in conformal field theory: deformation by a truly marginal operator.
{}From a spacetime point of view, this corresponds to a slow variation in the
VEV of a scalar field which has an exactly flat potential.\foot{To avoid
confusion, we remark that the present study focuses on static vacuum
solutions to string theory.  One expects that configurations involving
the generic slow variation
of such scalar fields are solutions as well.}
It is crucial to emphasize that these physically
smooth topology changing processes occur even at the level of
{\it classical}\/ string theory. It is not, as had been suspected
from point particle intuition, that quantum effects give rise to
topology change, but, rather, it is the extended structure of the string
which bears responsibility for this effect.

We can immediately summarize here the essential content of
\rW\ and \rAGM. From the viewpoint of classical general relativity or
the classical nonlinear sigma model, we know that there are
constraints on the metric tensor which appears in the action. Namely, since
the metric is used to measure lengths, areas, volumes, etc., it must
satisfy a set of positivity conditions. For instance, if we have a
nonlinear sigma model on a K\"ahler target space $M$ with metric
(in complex coordinates) $g_{\mu \overline \nu}$, we can write the
 K\"ahler form of the metric as
$J = ig_{\mu \overline \nu} dX^{\mu} \wedge dX^{\overline \nu}$ (a
real closed $2$-form).  The latter
must satisfy
\eqn\epositive{ \int_{M_r } J^r > 0}
where $M_r$ is an $r$ (complex) dimensional submanifold of $M$
and $J^r$ represents the $r$-fold wedge product of $J$ with itself.
The set of real closed $2$-forms which satisfy \epositive\ is a subset
of $H^2(X,\IR)$ known as the {\it K\"ahler}\/ cone and is schematically
depicted in figure 1a. Such K\"ahler forms manifestly span a cone
because if $J$ satisfies \epositive\ then so does $sJ$ for any positive
real $s$. The burden of \rW\ and \rAGM\ is that, in string theory,
\epositive\ can be relaxed and still result in perfectly well
behaved physics. In fact, the K\"ahler form of a target Calabi-Yau
space is one of the moduli fields of the associated conformal
field theory. Investigation of the conformal field theory moduli space
reveals that the corresponding geometrical description {\it necessarily}\/
involves configurations in which the (supposed) K\"ahler form lies outside of
the K\"ahler cone of the particular \CY\ being studied.
In fact, {\it any and all}\/ choices of an element of $H^2(X,\IR)$
give rise to well-defined conformal field theories.
In
\refs{\rW,\rAGM} it was shown that some of these configurations can be
interpreted as \NL s on \CY\ manifolds of topological type distinct
from the original. With respect to this \CY\ of new topology, the \K\
modulus satisfies \epositive\ and hence may be thought of as residing
in a new K\"ahler cone which shares a common wall with the original.
Furthermore, there is no physical obstruction to continuously deforming
the underlying conformal field theory so that its geometrical
description passes from one \K\ cone to another and hence results in
a change in topology of the target space -- i.e. of space itself.

\iffigs
\midinsert
$$
\matrix{\epsfxsize=2.5cm\epsfbox{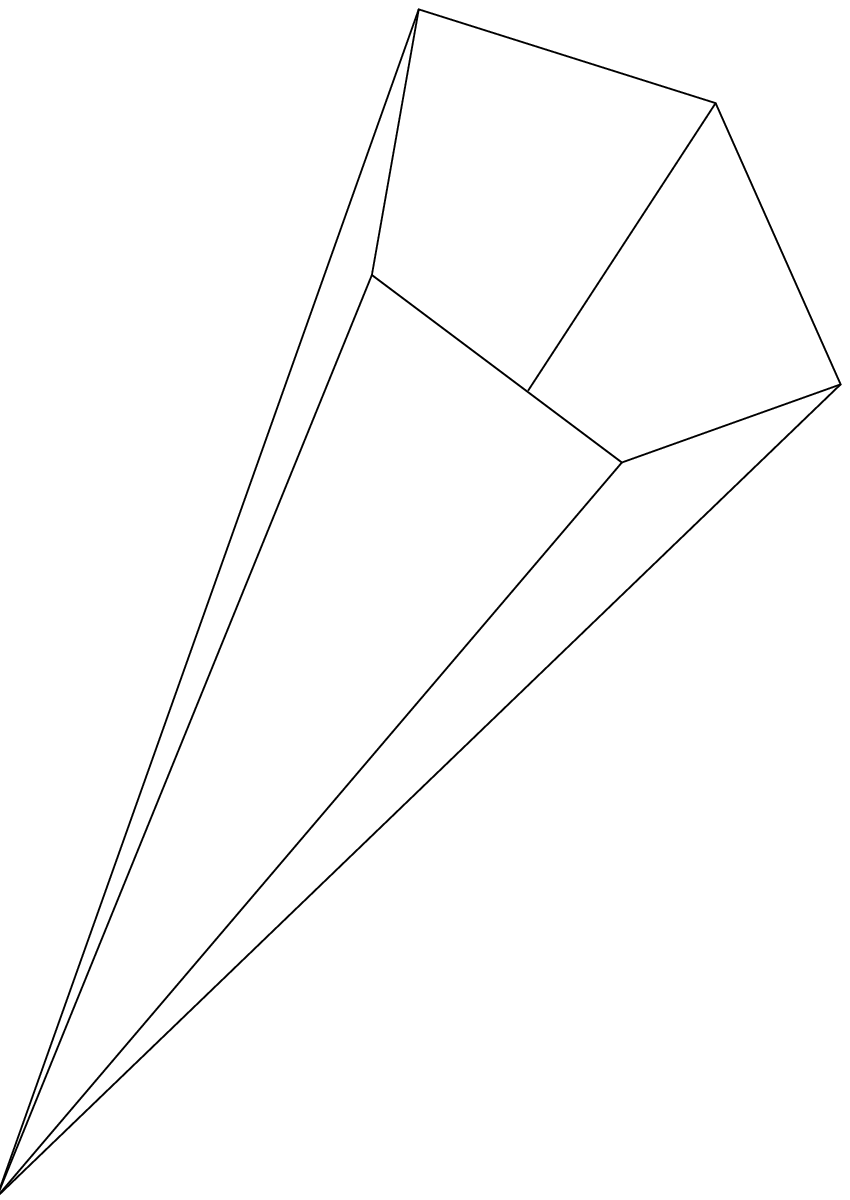} & \qquad &
\epsfxsize=2.5cm\epsfbox{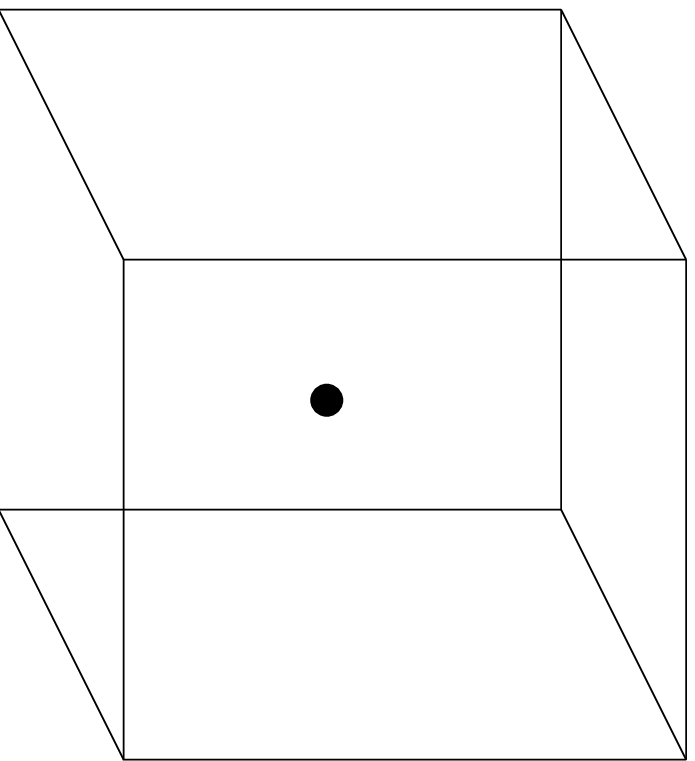} \cr
\quad & & \cr
\hbox{Figure 1a. K\"ahler cone.} & &
\hbox{Figure 1b. Domain of $w_l$'s.} \cr}
$$
\endinsert
\fi

These results were established in \rAGM\ by means of mirror symmetry.
Hence, in the next section we shall briefly review the phenomenon of
mirror manifolds. In section III we will give a discussion of
moduli spaces of both conformal theories and \CY\ manifolds in order
to fill in a bit more detail required for the discussion of
topology change. We will see that this discussion raises an interesting
puzzle whose resolution, discussed in section IV,
 directly leads to the necessity of physically
smooth topology changing processes. In section V we shall verify the
abstract discussion of the preceding sections in an explicit example
which provides a highly sensitive confirming test of the picture
we present. Finally, in section VI we shall give our conclusions.

\newsec{Mirror Manifolds}

Nonlinear sigma models on \CY\ target spaces, at their infrared
fixed point, provide the geometric interpretation for a class
of conformal field theories. As is well known, these conformal field
theories have $N = (2,2)$  world sheet supersymmetry. One can turn this
identification around and inquire as to
whether every $N = (2,2)$ superconformal
field theory with central charge, say, equal to $9$ is interpretable
as a nonlinear sigma model with some \CY\ as target. The answer to this
question is not known; however, it is known that one conformal field
theory can sometimes be interpretable in terms of nonlinear sigma
models on {\it two}\/ very different \CY\ target spaces
\ref\rGP{B.R. Greene and M.R. Plesser, Nucl. Phys. {\bf B338} (1990) 15.}.
The possible existence of this phenomenon was raised in
\ref\rDixon{L. Dixon, in {\it Superstrings, Unified Theories and
Cosmology 1987}\/ (G. Furlan et al., eds.), World Scientific, 1988, p. 67.}
and in \ref\rLVW{W. Lerche, C. Vafa and N. Warner, Nucl. Phys. {\bf B324}
(1989) 427.}
based on the fact that there is
an unnatural asymmetry between the identification of an abstract
conformal field theory with a \CY\ \NL\ which is resolved when a second
\CY\ interpretation exists. Namely, the truly marginal operators in
the abstract \CFT\ can be labeled with the $U(1)_{\rm L} \times U(1)_{
\rm R}$ quantum numbers of the lowest components of the supermultiplet
to which they belong. These eigenvalues divide the space of truly
marginal operators into those with charges $(1,1)$ and $(-1,1)$
(and their complex conjugates). Now, a \CY\ sigma model also has
two types of truly marginal operators: the complex structure deformations
and the \K\ deformations. Mathematically, these two types are vastly different
objects; nonetheless, since they correspond to the truly marginal
operators in the associated conformal field theory, the abstract
formulation only distinguishes them by the sign of a $U(1)$ charge.
It is surprising that an important mathematical distinction finds such
a trivial conformal field theory manifestation. It was suggested in
\rDixon\ and \rLVW\ that a natural resolution of this asymmetry would
be the existence of a second \CY\ manifold giving rise to the same
conformal field theory but with the identification of geometrical
deformations and conformal field theory marginal operators reversed
(relative to the $U(1)$ charges) with respect to the first \CY.
One consequence of the existence of such a second \CY\ interpretation
is that the Hodge numbers of the first, say $M$, and those of the second,
say $\tilde M$, are related via
\eqn\eHN{h^{p,q}_M = h^{3 - p,q}_{\tilde M} .}
Although an interesting speculation, there was no evidence for the
existence of this phenomenon until the simultaneous works of
\ref\rCLS{P. Candelas, M. Lynker and R. Schimmrigk, Nucl. Phys. {\bf B341}
(1990) 383.}\
and \rGP. The authors of \rCLS\ performed a computer
survey of a large number of \CY\ manifolds realized as hypersurfaces
in weighted projective four space. They found that
almost every \CY\ in the resulting list had a counterpart with
Hodge numbers related as above. This falls short of establishing
that these pairs of \CY\ manifolds
correspond to the same \CFT\ but it is at least consistent
with this possibility. In \rGP, on the other hand, a constructive proof
was given for the existence of certain pairs of Calabi-Yau manifolds
$M$ and $\tilde M$ whose Hodge numbers are related by
\eHN\ {\it and which give rise to isomorphic conformal field theories.}
Such pairs of \CY\ spaces were named ``mirror manifolds'' in \rGP\
because the relation between the Hodge numbers corresponds to a reflection
in a diagonal plane of the corresponding Hodge diamonds. The construction
of \rGP\ applies to any \CFT\ built up from the $N$=2  minimal models
and these include Fermat hypersurfaces in weighted projective space.
At the present time, this construction  supplies the only known examples
of mirror manifolds.\foot{There have been other conjectured
constructions of mirror manifolds in both the physics
\ref\rBH{P. Berglund and T. H\"ubsch, in
{\it Essays on Mirror Manifolds}, (S.-T. Yau, editor), International Press,
 1992, p. 388.}
and mathematics
\nref\rBat{V. Batyrev, ``Dual Polyhedra and Mirror Symmetry
for Calabi-Yau Hypersurfaces in Toric Varieties'', Essen preprint,
November 18, 1992.}
\nref\rBorisov{L. Borisov, ``Towards the Mirror Symmetry for Calabi-Yau
Complete intersections
in Gorenstein Toric Fano Varieties'', Univ. of Michigan preprint,
October, 1993.}%
\refs{\rBat,\rBorisov} literatures,
but as yet no one has been able to establish that these constructions
yield pairs of \CY\ manifolds corresponding to the same \CFT.}

Before proceeding, there are two points which, although not directly
relevant to our present study, are worth emphasizing here.
First, mirror manifolds are not the first nor the only examples of
distinct geometrical spaces which give rise to the same conformal field
theory. For example, string theory on a circle of radius $R$ and
on a distinct circle of radius $\alpha'/R$ give rise to isomorphic
physics \ref\rRoR{
K. Kikkawa, M. Yamasaki, Phys. Lett. {\bf 149B} (1984) 357\semi
N. Sakai, I. Senda, Prog. Theor. Phys. {\bf 75} (1986) 692.}.
The same is true for certain pairs of toroidal
orbifolds \ref\r{R. Dijkgraaf, E. Verlinde and H. Verlinde,
Commun. Math. Phys. {\bf 115} (1988) 649\semi
A. Shapere and F. Wilczek, Nucl. Phys. {\bf B230} (1989) 669.}.
The most general term describing the phenomenon
in quantum geometry of distinct spaces giving rise to identical physical
models is {\it string equivalent spaces}. That is, two distinct background
spaces $X$ and $Y$ on which string propagation is physically isomorphic
are called {\it string equivalent}. From this definition, it is clear
that $M$ and $\tilde M$ are a mirror pair if they are string equivalent
{\it and}\/ if they are \CY\ manifolds whose Hodge diamonds
are mirror reflected and hence related by \eHN.
In keeping with this definition, for instance,
circles of radii $R$ and $\alpha'/R$ are string equivalent but are
not a mirror pair.
Second, it has sometimes been asserted
that the phenomenon of mirror manifolds
amounts to nothing more than
the fact that there is a trivial automorphism
of $N = (2,2)$ conformal field theory obtained by changing the sign
of one of the $U(1)$ charges. There is a misleading
imprecision here. It is true that there is a trivial automorphism
of these conformal theories arising from such a change in sign.
However, this trivial conformal field theory operation has
an {\it equally trivial geometrical interpretation}. Namely, to specify
a supersymmetric nonlinear sigma model we need to supply not only a
\CY\ manifold but also a vector bundle on it
(to which the world sheet fermions couple), meeting certain
conditions. The simplest solution to these conditions and the solution
implicitly chosen in most studies is that of the tangent bundle to
the \CY\ manifold. There is, however another equally valid and physically
equivalent choice: the {\it cotangent}\/ bundle
to the same \CY\ space. These two equivalent
choices differ, from the conformal field theory viewpoint, by
a change in sign of one of the $U(1)$ charges in the theory. Thus,
as promised, a trivial \CFT\ operation has a trivial geometric
interpretation. {\it This is not mirror symmetry.} Rather, mirror symmetry
is a phenomenon in which the space changes, not simply the bundle.
It is true that this  isomorphism
is proved \rGP\ by making use of  the fact that the
two relevant conformal theories differ by the trivial automorphism
associated with the $U(1)$ charge. However, the existence of
such a trivial automorphism does not (at our present level of
understanding) by any means establish the existence of mirror manifolds ---
in fact, as just mentioned, there is a far more trivial geometric
interpretation which immediately presents itself.

There are a number of interesting and important implications of
mirror symmetry which we will not have time to discuss here. However,
one particular result will be useful in our later discussion, so
we briefly record it now.

Since a mirror pair $M$ and $\tilde M$
correspond to the same conformal field theory,
every correlation function in the latter has two geometric interpretations:
one on $M$ and one on $\tilde M$. Typically, these geometric realizations
will be quite different; however, since they mathematically represent
one and the same correlation function they must be identically equal.
This fact gives rise to some highly nontrivial identities between
particular geometrical formulas on $M$ and others on $\tilde M$.
One such identity, originally shown in \rGP\ and later employed to
remarkable ends in \ref\rCDGP{P. Candelas, X.C. de la Ossa, P.S. Green,
 and L. Parkes,
Phys. Lett. {\bf 258B} (1991) 118; Nucl. Phys. {\bf B359} (1991) 21.},
arises from the study of three point functions
amongst fields $\{O_i\}$ associated with, say, the $(\rm chiral, \rm chiral)$
primary fields in the conformal theory. On $\tilde M$
such fields are associated
with harmonic $(0,1)$ forms taking values in the
tangent bundle, $B_{(i)}^\alpha$, and it has been shown that
\ref\rStromWitten{A. Strominger and E. Witten, Commun. Math. Phys. {\bf 101}
(1985) 341.}
\nref\rStrom{A. Strominger, Phys. Rev. Lett. {\bf 55} (1985) 2547.}%
\nref\rDSWW{M. Dine, N. Seiberg, X.-G. Wen, and E. Witten, Nucl. Phys.
{\bf B278} (1987) 769; Nucl. Phys. {\bf B289} (1987) 319.}%
\nref\rAM{P.S. Aspinwall and D.R. Morrison, Commun. Math. Phys. {\bf 151}
(1993) 245.}%
\eqn\eto{\langle O_i O_j O_k\rangle = \int_{\tilde M} \Omega
\wedge B_{(i)}^\alpha \wedge B_{(j)}^\beta \wedge B_{(k)}^\gamma
\Omega_{\alpha \beta \gamma} .}
On $M$, the $O_i$ are associated with harmonic $(0,1)$ forms taking
values in the cotangent bundle which are isomorphic to harmonic
$(1,1)$ forms $A_{(i)}$ and it has been shown that
\refs{\rStrom,\rDSWW,\rCDGP,\rAM}
\eqn\eti{\eqalign{
\langle O_i O_j O_k\rangle
	=& \int_M A_{(i)} \wedge A_{(j)} \wedge A_{(k)}+\cr
	\qquad&
\sum_{m,\{u \}}\ex{\int_{\CP1 }u_m^*K}
  \left(\int_{\CP1 } u^*A_{(i)}   \int_{\CP1 } u^*A_{(j)}
 \int_{\CP1 } u^*A_{(k)} \right),\cr}}
where
$\{u \}$ is the set of holomorphic maps to rational curves on $M$,
$u: \CP1 \rightarrow C $ (with $C$ such  a holomorphic  curve),
$\pi_m$ is an $m$-fold cover $\CP1 \rightarrow \CP1$ and
$u_m = u \circ \pi_m$.

As each of these mathematical expressions on the right hand side is
equal to the same correlation function in a single conformal field theory,
they must be equal to each other. Hence we have
\eqnn\eequal
$$\displaylines{
\int_{\tilde M} \Omega
\wedge B_{(i)}^\alpha \wedge B_{(j)}^\beta \wedge B_{(k)}^\gamma
\Omega_{\alpha \beta \gamma}
= \hfill\eequal \cr
\int_M A_{(i)} \wedge A_{(j)} \wedge A_{(k)} +
\sum_{m,\{u \}}\ex{\int_{\CP1 }u_m^*K}
  \left(\int_{\CP1 } u^*A_{(i)}   \int_{\CP1 } u^*A_{(j)}
 \int_{\CP1 } u^*A_{(k)} \right) .}$$
In fact, although for ease of discussion we have focused on a single
conformal field theory, if we deform that theory to any point in its
moduli space, there exist corresponding choices for the K\"ahler class
on $M$ and the complex structure of $\tilde M$ such that this equality
continues to hold.
Notice that the leading term in \eto\ is the topological intersection
form on $M$ and that this term is the only
one which contributes to the correlation function
in \eto\ if the integral $\int_C K$ goes to infinity, for every rational
curve $C$ on $M$.
This occurs if the K\"ahler form $K$ approaches a ``large radius limit''
--- a concept which will be made precise in the sequel.
\newsec{Moduli Spaces}

Quite generally, as mentioned, the conformal field theories
we study here
come in continuously connected families related via
deformations by truly marginal operators. For the $N$=2 theories,
more specifically,
these truly marginal operators come in two varieties which
are distinguished by their $U(1) \times U(1)$ charges with the latter
being a subalgebra of the $N$=2 superconformal algebra. In particular,
the two types of marginal operators have charges $(1,1)$ and
$(-1,1)$ respectively. (Actually, the marginal operators are chargeless ---
they lie in supermultiplets whose lowest component has the given
charges.) Invoking standard usage, we refer to the space of all conformal
theories related by such truly marginal deformations as the
{\it conformal theory moduli space}.

When an $N$=2 conformal theory arises from a nonlinear sigma model
with a Calabi-Yau target space, the marginal operators just referred to
have geometrical counterparts. The two types of marginal operators
correspond to the two types of deformations of the Calabi-Yau space
which preserve the Calabi-Yau condition (of Ricci flatness). These
are deformations of the complex structure and deformations of the
K\"ahler structure. Concretely, one can think of the latter as Ricci flat
deformations of the metric of the form $\delta g_{\mu \overline \nu}$
and the while the former are of the form $\delta g_{\mu \nu}$.

As our analysis will involve a close study of these moduli spaces,
let us now describe each in a bit more detail.

\subsec{K\"ahler Moduli Space}

Given a K\"ahler metric $g_{\mu \overline \nu}$ we can construct the
K\"ahler form
$J = ig_{\mu \overline \nu} dX^{\mu} \wedge dX^{\overline \nu}$.
As discussed earlier, the set of allowed $J$'s forms a cone
known as the K\"ahler cone of $M$. One additional important fact is that
string theory instructs us to work not just with $J$ but also with
$B = B_{\mu \overline \nu}$ the antisymmetric tensor field. The latter,
which is a closed two-form,
combines with $J$ in the form $B + iJ$ to yield
 the highest component of a complex chiral
multiplet we shall call $K$. $K$ can therefore be thought of as a
{\it complexified}\/ K\"ahler form. The precise way in which $B$ enters
the conformal field theory is such that if $B$ is replaced by $B + Q$
with $Q \in H^2(M,\BZ)$, then the resulting physical model does not
change. Thus, a convenient way to parametrize the space of allowed
and physically distinct $K$'s is to introduce
\eqn\ew{ w_l = e^{2 \pi i (B_l + iJ_l)} }
where we have expressed
\eqn\eBJ{ B + iJ = \sum_l (B_l + iJ_l) e^l}
with the $e^l$ forming an integral basis for $H^2(M,\BZ)$.
The $w_l$ have the invariance of the antisymmetric tensor field under
integral shifts built in; the constraint that $J$ lie in the
K\"ahler cone bounds the norm of the $w_l$. Thus, the K\"ahler cone
and space of allowed and distinct $w_l$ are schematically shown in
figures 1a and 1b. Notice that any choice of complexified K\"ahler form
in the interior of figure 1b is physically admissible. Choices of
$K$ which correspond to points on the walls in figure 1b (or 1a)
correspond to metrics on $M$ which fail to meet \epositive\ and hence are
degenerate in some manner.

\subsec{Complex Structure Moduli Space}

All of the Calabi-Yau spaces we shall concern ourselves with here
are given by the vanishing locus of homogeneous polynomial constraints
in some projective space (or possibly a weighted projective space and
products thereof). For ease of discussion, and in preparation for an
explicit example we will examine shortly, let's assume we are dealing with
a Calabi-Yau manifold given by the vanishing locus of a homogeneous
polynomial  $P$ of degree $d$ in weighted projective four space
$\CP4_{\{k_1,\ldots,k_5\}}$. The Calabi-Yau condition translates into the
requirement that $d = \sum_i k_i$. Let's call the homogeneous
weighted projective space coordinates $(z_1,\ldots,z_5)$ and write down
the most general form for $P$:
\eqn\eP{ P = \sum a_{i_1 i_2 \ldots i_5} z_1^{i_1}\ldots z_5^{i_5} }
where $\sum_j k_j i_j = d$.
Different choices for the constants $a_{i_1 i_2 \ldots i_5}$
correspond to different choices for the complex structure of the
underlying Calabi-Yau manifold. There are two important points
worthy of emphasis in this regard. First, not all choices of the
$a_{i_1 i_2 \ldots i_5}$ give rise to distinct complex structures.
For instance, distinct choices of the $a_{i_1 i_2 \ldots i_5}$ which can
be related by a rescaling of the $z_j$ of the form
$z_j \rightarrow \lambda_j z_j$ with $\lambda_j \in \IC^*$ manifestly
correspond to the same complex structure (as they differ only by a trivial
coordinate transformation). The most general situation would require
that we consider $a_{i_1 i_2 \ldots i_5}$'s related by general linear
transformations on the $z_j$'s. Second, not all choices of
$a_{i_1 i_2 \ldots i_5}$ give rise to smooth Calabi-Yau manifolds.
Specifically, if the $a_{i_1 i_2 \ldots i_5}$ are such that $P$ and
$\partial P \over \partial z_j$ have a common zero (for all $j$),
then the space given by the vanishing locus of $P$ is not smooth.
The set of all choices of the coefficients $a_{i_1 i_2 \ldots i_5}$ which
correspond to such singular spaces comprise the {\it discriminant locus}\/
of the family of Calabi-Yau spaces associated with $P$. The precise
equation of the discriminant locus is generally quite complicated;
however, the only fact we need is that it forms a complex codimension one
subspace of the complex structure moduli space.
{}From the viewpoint of conformal field theory, the nonlinear sigma
model associated to points on the discriminant locus appears to
be ill defined. For example, the chiral ring becomes infinite
dimensional. It is an interesting and important question to thoroughly
understand whether there might be some way of making sense of such
theories. For the present purposes, though, all we need to know is
that at worst the space of badly behaved physical models is complex
codimension one in the complex structure moduli space. We illustrate
the form of the complex structure moduli space in figure 2.

\iffigs
\midinsert
$$\vbox{\centerline{\epsfxsize=10cm\epsfbox{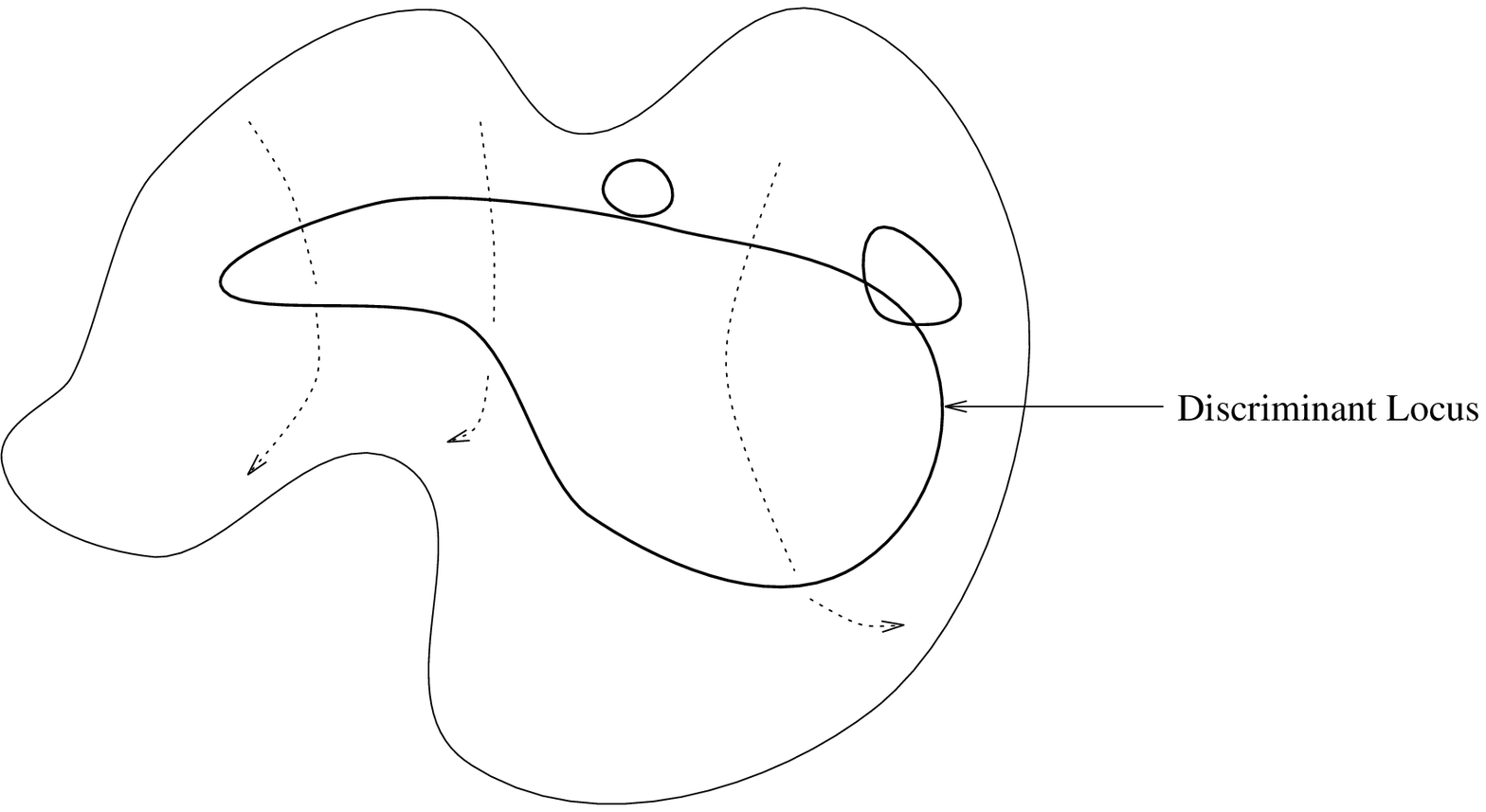}}
\centerline{Figure 2. The moduli space of complex structures.}}$$
\endinsert
\fi

\subsec{Implications of Mirror  Manifolds}

Locally the moduli space of Calabi-Yau deformations is a product space
of the complex and K\"ahler deformations (in fact, up to subtleties which
will not be relevant here, we can think of the moduli space as a global
product).  Thus,  we expect
\eqn\eMS{ {\cal M}_{\rm CFT} \equiv {\cal M}_{\hbox{\ninepoint
complex structure}}
\times {\cal M}_{\hbox{\ninepoint K\"ahler structure}} }
with ${\cal M}_{(\ldots)}$ denoting the moduli space of $(\ldots)$.
Pictorially, we can paraphrase this by saying that the conformal
field theory moduli space is expected to be
the product of figure 1b and figure 2.

This, in fact, was the picture which had emerged from much
work over the last few years and was generally accepted.
The advent of mirror symmetry, however, raised a serious puzzle related to
this description (as first observed in
\ref\rAL{P.S. Aspinwall and C.A. L\"utken,
Nucl. Phys. {\bf B355} (1991) 482.}).
Let $M$ and $\tilde M$ be a mirror pair of
Calabi-Yau spaces. As we discussed before, such a pair correspond
to isomorphic conformal theories with the explicit isomorphism
being a change in  sign of, say, the right moving $U(1)$
charge. From our description of the moduli space, it then follows that
the moduli space of K\"ahler structures on $M$ should be isomorphic to
the moduli space of complex structures on $\tilde M$
and vice versa. That is,
both $M$ and $\tilde M$ correspond to the same family of conformal
theories and hence yield the same moduli space on the left hand
side of \eMS. Therefore, the right hand side of \eMS\ must also be
the same for both $M$ and $\tilde M$. The explicit isomorphism of
mirror symmetry shows this to be true with the two factors on
the right hand side of \eMS\ being interchanged for $M$ relative
to $\tilde M$.

The isomorphism of the K\"ahler moduli space of one Calabi-Yau and the
complex structure of its mirror is a statement which appears to
be in direct conflict with the form of figure 1b and that of figure 2.
Namely, the former is a bounded domain while the latter is a quasi-projective
variety. More concretely, the subspace of theories which appear possibly to
be badly behaved are the boundary points in figure 1b (where the metric
on the associated Calabi-Yau fails to meet \epositive) and the points
on the discriminant locus in figure 2. The former are real codimension 1
while the latter are real codimension 2. Therefore, how can these two
spaces be isomorphic as implied by mirror symmetry?

\newsec{Topology Change}

As the puzzle raised in the last section was phrased in terms  of
those points in the moduli space which have the potential to
correspond to badly behaved theories, it proves worthwhile to study
the nature of such points in more detail. We will first do this from
the point of view of the K\"ahler moduli space of $M$.

Consider a path in the K\"ahler moduli space which begins deep in
the interior and moves towards and finally reaches a boundary wall
as illustrated in figure 3. More specifically, we follow a path
in which the area of a $\CP1$ (a rational curve) on $M$ is continuously
shrunk down to zero, attaining the latter value on the wall itself.
The question we ask ourselves is: does this choice for the K\"ahler form
on $M$ yield an ill defined conformal theory and furthermore, what would
happen if we try to extend our path beyond the wall where it appears
that the area of the rational curve would become negative?
(We note the linguistically awkward phrase ``area of a curve'' arises
since we are dealing with complex curves which therefore are real
dimension two.)

\iffigs
\midinsert
$$\vbox{\centerline{\epsfxsize=4cm\epsfbox{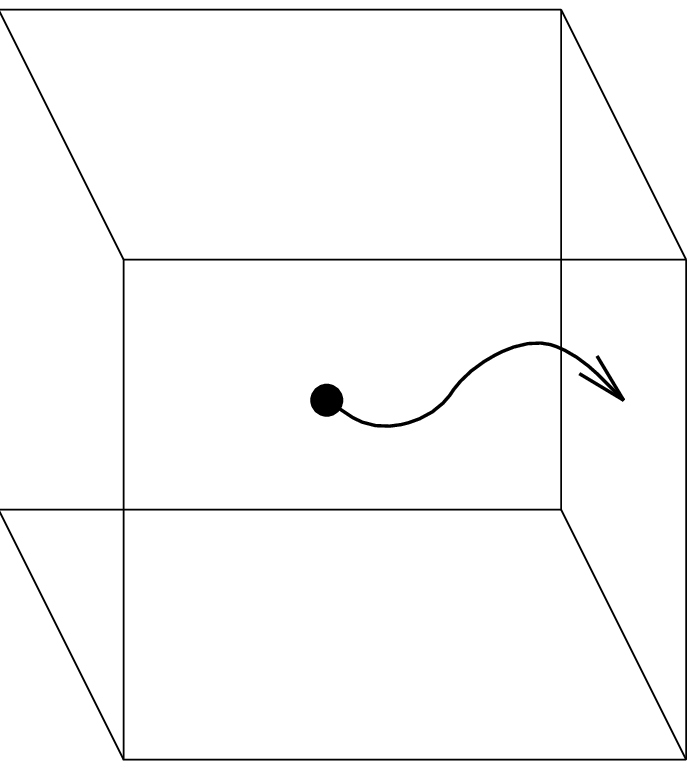}}
\centerline{Figure 3. A path to the wall.}}$$
\endinsert
\fi

As a prelude to answering this physical question,
we note that precisely this operation is well known and thoroughly
studied from the viewpoint of mathematics. Namely, in algebraic
geometry there is an operation called a {\it flop}\/ in which the area of
a rational curve is shrunk down to zero ({\it blown down}) and then
expanded back to positive volume ({\it blown up}) in a ``transverse''
direction.
Typically (although not always) this operation results in a change of
the topology of the space in which the curve is embedded. Thus, when
we say that the blown up curve has positive volume we mean positive
with respect to the K\"ahler metric on the new ambient space.
That is, the flop operation involves first following a path like
that in figure 3 which blows the curve down, and then continuing through
the wall (as in figure 4) by blowing the curve up to positive volume on a new
Calabi-Yau space. The latter space, $M'$ also has a K\"ahler cone whose
complexification in the exponentiated $w_l$ coordinates is another
bounded domain. Thus, the operation of the flop corresponds to a path
in moduli space beginning in the K\"ahler cone of $M$, passing through
one of its walls and landing in the
{\it adjoining}\/ K\"ahler cone of $M'$. Although $M$ and $M'$ can be
topologically distinct, their Hodge numbers are the same; they differ
in more subtle topological invariants such as the intersection form
governing the classical homology ring. Mathematically, they are said
to be topologically distinct but in the same birational equivalence class.

\iffigs
\midinsert
$$\vbox{\centerline{\epsfxsize=8cm\epsfbox{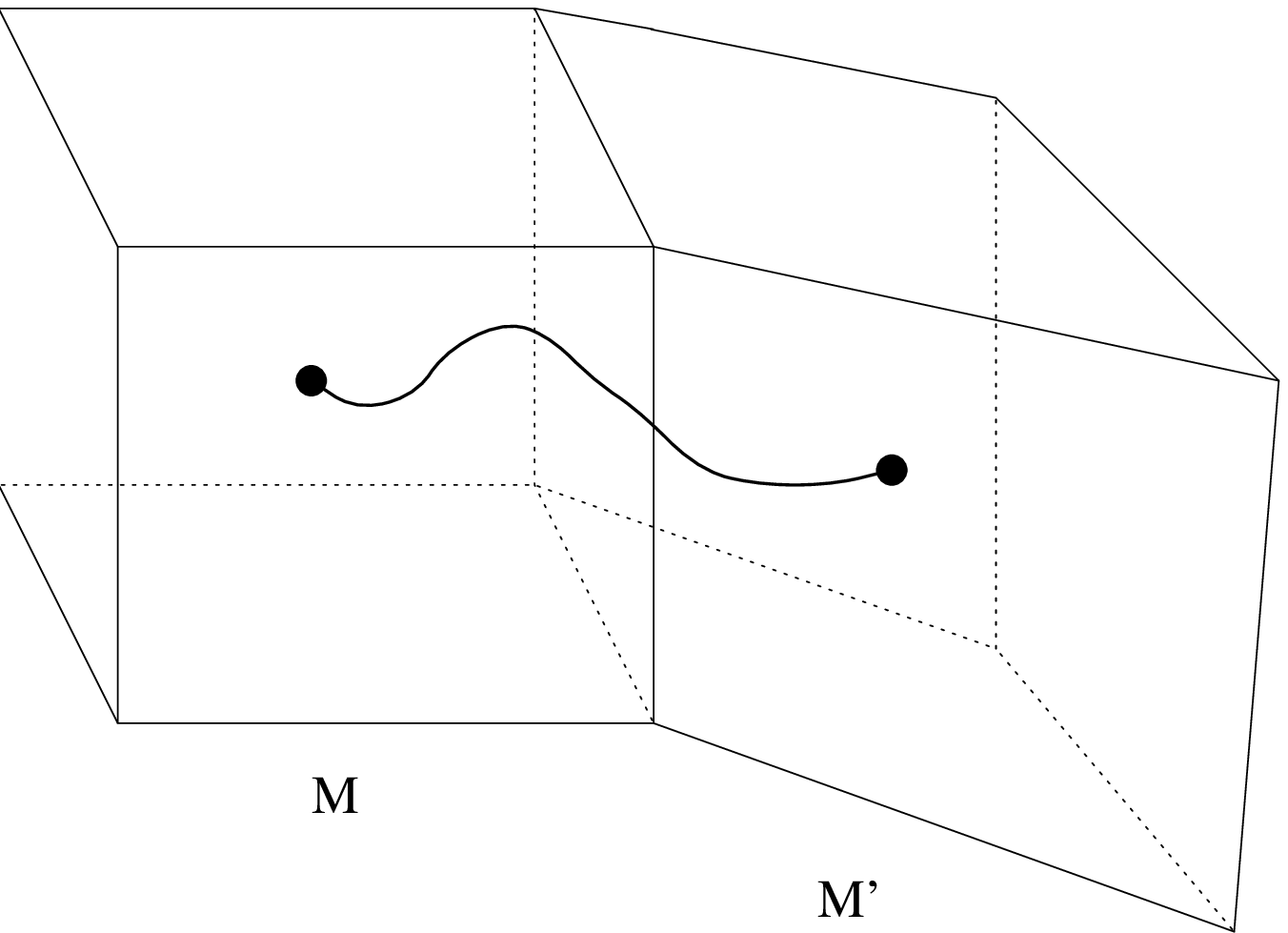}}
\centerline{Figure 4. A topology-changing path.}}$$
\endinsert
\fi

The mathematical formulation of what it means to pass to a wall in the
K\"ahler moduli space has led us to a more detailed framework for
studying the corresponding description in conformal field theory. We
see that from the mathematical point of view, distinct K\"ahler moduli
spaces naturally adjoin along common walls. We can rephrase our
initial motivating question of two paragraphs ago as: does the operation
of flopping a rational curve (and thereby changing the  topology
of the Calabi-Yau under study) have a physical manifestation? That is,
does a path such as that in figure 4 correspond to a family of well
behaved conformal theories?

This is a hard question to answer directly because our main tool
for analyzing nonlinear sigma models is perturbation theory. The
expansion parameters of such perturbative studies are
of the form ${\sqrt \alpha'}/R$
where $R$ refers to the set of K\"ahler moduli on the target manifold.
Now, when we approach or reach a wall in the K\"ahler moduli space,
at least one such moduli field $R$ is going to zero
(namely the one which sets the size of the blown down
rational curve). Hence, sigma model
perturbation theory breaks down and we are hard pressed to answer
directly whether
the associated conformal theory makes nonperturbative sense.

This situation --- one in which we require a nonperturbative
understanding of observables on $M$ --- is tailor made for an analysis
based upon mirror symmetry. Perturbation theory breaks down on $M$
because of the degenerate (or nearly degenerate) choice of its
K\"ahler structure. Note that all of our discussion could be carried
through for any convenient (smooth)
 choice of its complex structure.  Via mirror
symmetry, this implies that the relevant analysis for answering the
question raised two paragraphs ago should be carried out on $
\tilde M$ for
a particular form of the {\it complex structure}\/ (namely, that which
is mirror to the degenerate K\"ahler structure on $M$) but for any
convenient choice of the K\"ahler structure. The latter, though,
determines the applicability of sigma model perturbation theory on
$\tilde M$. Thus, we can choose this K\"ahler structure to be
arbitrarily ``large'' (that is, distant from any walls in the K\"ahler cone)
and hence arrange things so that we can completely trust perturbative
reasoning. In other  words, by using mirror symmetry we have rephrased
the difficult and necessarily nonperturbative question of whether conformal
field theory continues to make sense for degenerating K\"ahler structures
in terms of a purely perturbative question on the mirror manifold.

This latter perturbative question is one which is easy to answer and, in
fact, we have already done so in our discussion of the complex structure
moduli space. For large values of the K\"ahler structure (again, this
simply means that we are far from the walls of the K\"ahler cone)
the only choices of the complex structure which yield (possibly)
badly behaved conformal theories are those which lie on the discriminant
locus. As noted earlier, the discriminant locus is complex codimension
one in the moduli space (real codimension two). Thus, the complex structure
moduli space is, in particular, path connected. Any two points can be
joined by a path which only passes through well behaved theories;
in fact, the generic path in the complex structure moduli space has
the latter property.
This is the answer to our question.
By mirror symmetry, this conclusion must hold for
a generic path in  K\"ahler moduli space and hence it would seem that
a topology changing path such as that of figure 4 (by a suitable small
jiggle at worst) is a physically well behaved process.
Even though the metric degenerates, the physics of string theory
continues to make sense. We are already familiar from the foundational
work on orbifolds \ref\rDHVW{L. Dixon, J. Harvey, C. Vafa and E. Witten,
Nucl. Phys.
{\bf B261} (1985) 678; Nucl. Phys. {\bf B274} (1986) 285.}\
that degenerate metrics
can lead to sensible string physics. Now we see that  physically
sensible degenerations of other types (associated to flops)
can alter the topology of the universe.
In fact,
the operation being described
 --- deformation by a truly marginal operator --- is amongst
the most basic and common physical processes in conformal field theory.

To summarize the picture of moduli space  which has emerged from this
discussion we refer to figure 5. The conformal field theory moduli space
is geometrically interpretable in terms of the product of a complex
structure moduli space and an  enlarged K\"ahler moduli
space ${\cal M}_{\hbox{\ninepoint enlarged K\"ahler}}$.
The latter contains  numerous complexified K\"ahler cones
of birationally equivalent yet topologically distinct
Calabi-Yau manifolds adjoined along common walls.\foot{The union
of such regions constitutes what we call the ``partially enlarged''
K\"ahler moduli space.  The enlarged K\"ahler moduli space
includes additional regions as we shall mention shortly.}
 There are two such
geometric interpretations, via mirror symmetry, with the roles of complex
structure and K\"ahler structure being interchanged. This is also
indicated in figure 5.

\iffigs
\midinsert
$$\vbox{\centerline{\epsfxsize=4.25in\epsfbox{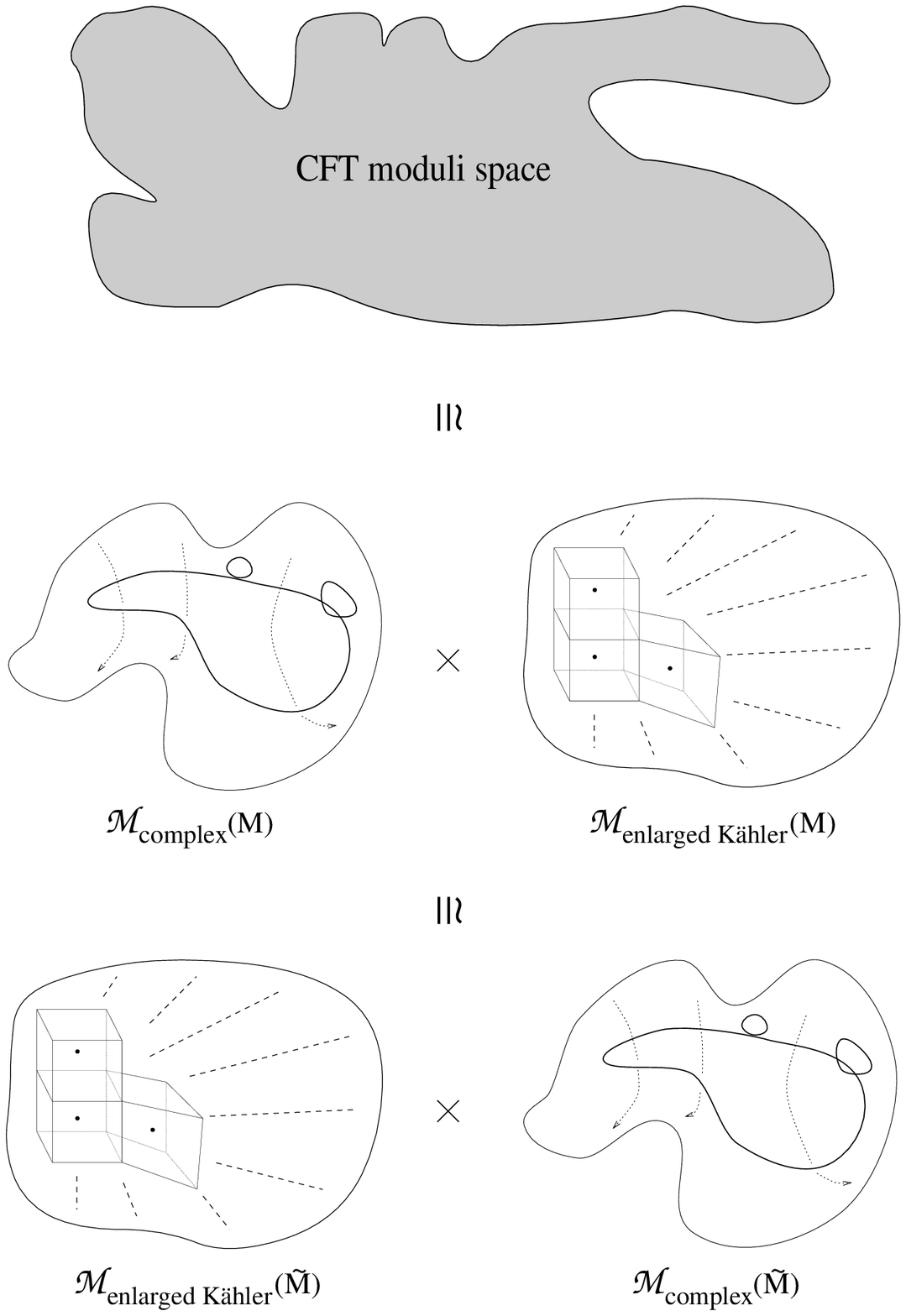}}\vskip.5cm
\centerline{Figure 5. The conformal field theory moduli space.}}$$
\endinsert
\fi

We should stress that from an abstract point of view this is a compelling
picture. Although we do not have time or space to discuss it here,
the augmentation of the K\"ahler moduli space in the manner presented
(and, more precisely, as we will generalize shortly) gives it
a mathematical structure which is {\it identical}\/ to that of the
complex structure moduli space of its mirror. In the important case of
Calabi-Yau's which are toric hypersurfaces, both of these moduli
spaces are realized as
identical compact {\it toric varieties}. Hence, the picture presented resolves
the previous troubling asymmetry between the structure of these
two spaces which are predicted to be isomorphic by mirror symmetry.

Although compelling, we have not proven that the picture
we are presenting is correct. We have found a natural mathematical
structure in algebraic geometry which {\it if}\/ realized by the
physics of conformal field theory resolves some thorny issues in
mirror symmetry. We have not established, as yet, whether conformal field
theory makes use of this compelling mathematical structure.
If conformal field theory does avail itself of this structure, though,
there is a very precise and concrete conclusion we can draw: every point
in the (partially) enlarged K\"ahler moduli space of $M$ must correspond
under mirror symmetry to some point in the complex structure moduli space
of $\tilde M$. This implies, of course, that any and all observables
calculated in the theories associated to these corresponding points
must be identically equal. Let's concentrate on the three point functions
we introduced earlier in \eequal. As we discussed, if we choose a point
in the K\"ahler moduli space for which the instanton corrections are
suppressed, the correlation function approaches the topological intersection
form on the Calabi-Yau manifold.
For ease of calculation, we shall study the correlation functions of
\eti\ in this limit.
This analysis will be similar to that presented in
\ref\rALR{P.S. Aspinwall, C.A. L\"utken, and G.G. Ross, Phys. Lett. {\bf 241B}
(1990) 373.}
although in this case in the (partially) enlarged
K\"ahler moduli space, there is not a single unique ``large radius'' point
of the sort we are looking for.
Rather, every cell in the (partially) enlarged moduli
space supplies us with one such point. Since these cells are the complexified
K\"ahler cones of topologically distinct spaces, the intersection forms
associated with these large radius points are different. If the
moduli space picture we are presenting in figure  5 is correct, then
there must be points in the complex structure moduli space of the mirror
whose correlation functions exactly reproduce  each and every one of
these intersection forms.
This is a precise and concrete statement whose veracity would provide
a strong verification of the picture presented in
figure 5.
In the next section we carry out this verification in a particular
example.

\newsec{An Example}

In this section we briefly  carry out the abstract program discussed in
the last few sections in a specific example. We will see that the
delicate predictions just discussed can be explicitly verified.

We focus on the Calabi-Yau manifold  $M$ given by the vanishing locus
of a degree $18$ homogeneous polynomial in the weighted projective
space $\CP4_{\{6,6,3,2,1\}}$ and its mirror $\tilde M$. For the former we can
take the polynomial constraint to be
\eqn\eeightteen{
z_0^3+z_1^3+z_2^6+z_3^9+z_4^{18}
+a_0 z_0z_1z_2z_3z_4 = 0 }
where the $z_i$ are the homogeneous weighted space coordinates
and $a_0$ is a large and positive constant (whose value, in fact, is
inconsequential to the calculations which follow). The mirror to this
family of Calabi-Yau spaces is constructed via the method of
\rGP\ by taking an orbifold of $M$ by the maximal scaling symmetry
group $\IZ_3 \times \IZ_3 \times \IZ_3$.

A study of the K\"ahler structure of $M$ reveals that there are
five cells in its (partially) enlarged K\"ahler moduli space,
each corresponding to a sigma model on a smooth topologically distinct
Calabi-Yau manifold. In each of these cells there is a large radius point
for which instanton corrections are suppressed  and hence the correlation
functions of \eti\ are just the intersection numbers of the respective
Calabi-Yau's. We have calculated these for each of the five birationally
equivalent yet topologically distinct Calabi-Yau spaces and we record
the results in  table 1. To avoid having to deal with issues associated
with normalizing fields in the subsequent discussion, in table 1 we
have chosen to list our results in terms of ratios of correlation functions
for which such normalizations are irrelevant.
(The $D_i$ and $H$ are divisors on $M$, corresponding to  elements
in $H^1(M,T^*)$ by Poincar\'e duality.)

\midinsert
\def\tablerule{\noalign{\hrule}}
\def\ilspace{\omit&height2pt&&&&&&&&&&&&\cr}
$$\vbox{\tabskip=0pt \offinterlineskip
\halign{\strut#&
\vrule#&\hfil\quad$#$\quad\hfil&\vrule#&\hfil\quad$#$\quad\hfil&
\vrule#&\hfil\quad$#$\quad\hfil&\vrule#&\hfil\quad$#$\quad\hfil&
\vrule#&\hfil\quad$#$\quad\hfil&\vrule#&\hfil\quad$#$\quad\hfil&
\vrule#\cr \tablerule\ilspace
&&\hbox{Resolution}&&\Delta_1&&\Delta_2&&\Delta_3
&&\Delta_4&&\Delta_5&\cr\ilspace\tablerule\ilspace
&&{(D_1^3)(D_4^3)\over(D_1^2D_4)(D_1D_4^2)}
&& -7 && 0/0 && 0/0 && \infty && 9 &\cr\ilspace
&&{(D_2^2D_4)(D_3^2D_4)\over(D_2D_3D_4)(D_2D_3D_4)}
&& 2 && 4 && 0 && 0/0 && 0/0 &\cr\ilspace
&&{(D_2D_3D_4)(HD_2^2)\over(D_2^2D_4)(HD_2D_3)}
&& 1 && 1 && 1 && 0 && 0/0 &\cr\ilspace
&&{(D_2D_3D_4)(HD_1^2)\over(D_1^2D_4)(HD_2D_3)}
&& 2 && 1 && \infty && 0/0 && 0 &\cr\ilspace
\tablerule
\noalign{\vskip2pt}
&\multispan{13}\hfil Table 1: Ratios of intersection numbers\hfil\cr
}}$$
\endinsert

Following the discussion of the last section, our goal now is to
find five limit points  in the complex structure moduli space of $\tilde M$
such that appropriate ratios of correlation functions yield the
same results as in table 1. To do so, we note that the most general
complex structure on $\tilde M$ can be written
\eqn\ecsmirror{\eqalign{W=z_0^3&+z_1^3+z_2^6+z_3^9+z_4^{18}\cr
&+a_0 z_0z_1z_2z_3z_4 +
a_1 z_2^3z_4^9 + a_2 z_3^6z_4^6 + a_3 z_3^3z_4^{12} +
a_4 z_2^3z_3^3z_4^3 =0.\cr}}
We will describe these limit points by parametrizing the complex structure
as $a_i = s^{r_i}$ for real parameters $s$ and $r_i$ and we send $s$ to
infinity. The  limit points are therefore distinguished by the {\it rates}\/
at which the $a_i$ approach infinity. Our task, therefore, is to find
appropriate values for the $r_i$ (if they exist) such that we obtain mirrors
to the five large radius Calabi-Yau spaces of the last paragraph.
The technique we use to do this is to describe both the complex
structure moduli space of $\tilde M$ and the enlarged K\"ahler moduli
space of $M$ in terms of toric geometry. This description, at a fundamental
level, makes it manifest that these two moduli spaces are isomorphic.
We do not have time to present such analysis here --- rather, we refer
the reader to \rAGM. For the present purpose we note that
a direct outcome of this analysis is a prediction for five choices of
the vector $(r_0,\ldots,r_4)$ which should yield the desired mirrors.
As we have discussed, a sensitive test of these predictions is to
calculate the mirror of the ratios of correlation functions in table 1
(using \eto\
and the method of
\ref\rPC{P. Candelas, Nucl. Phys. {\bf B298} (1988) 458.})
for each of these complex
structure limits and see if we get the same answers. We have done this
and we show the results in table 2. Note that in the limit
$s$ goes to infinity we get precisely the same results.
(The $\varphi_i$ are elements of $H^1(\tilde M,T)$.)

\mark{B}\setbox\Bbox=\vbox{
{\ifx\answ\bigans\eightpoint\fi\hfuzz=10cm
\def\doots{\hbox to 4truept{$\dots$}}
\def\lquad{\hskip.3em\relax}
\def\tablerule{\noalign{\hrule}}
\def\ilspace{\omit&height2pt&&&&&&&&&&&&\cr}
$$\vbox{\tabskip=0pt \offinterlineskip
\halign{\strut#&
\vrule#&\hfil\quad$#$\quad\hfil&\vrule#&\hfil\lquad$#$\quad\hfil&
\vrule#&\hfil\lquad$#$\quad\hfil&\vrule#&\hfil\lquad$#$\quad\hfil&
\vrule#&\hfil\lquad$#$\quad\hfil&\vrule#&\hfil\lquad$#$\quad\hfil&
\vrule#\cr \tablerule\ilspace
&&\hbox{Resolution}&&\Delta_1&&\Delta_2&&\Delta_3&&
      \Delta_4&&\Delta_5&\cr\ilspace\tablerule\ilspace
&&\hbox{Direction}&&\omit\hfil$
({11\over9},1,{4\over3},{5\over3},{5\over3})
$\hfil&&\omit\hfil$
({7\over6},{1\over2},1,1,{5\over2})
$\hfil&&\omit\hfil$
({3\over2},{1\over2},2,3,{3\over2})
$\hfil&&\omit\hfil$
({13\over9},2,{5\over3},{4\over3},{4\over3})
$\hfil&&\omit\hfil$
({11\over6},{7\over2},1,1,{3\over2})
$\hfil&\cr\ilspace\tablerule\ilspace
&&{\langle \varphi_1^3\rangle\langle \varphi_4^3\rangle\over\langle
\varphi_1^2\varphi_4\rangle\langle \varphi_1\varphi_4^2\rangle}
&& -7-181s^{-1}+\doots &&  &&  && -{2\over5}s^2-{129\over250}s+\doots
&& 9+289s^{-1}+\doots &\cr\ilspace
&&{\langle \varphi_2^2\varphi_4\rangle\langle \varphi_3^2\varphi_4
\rangle\over\langle \varphi_2\varphi_3\varphi_4\rangle\langle
\varphi_2\varphi_3\varphi_4\rangle}
&& 2-5s^{-1}+\doots && 4-22s^{-1}+\doots && 0+2s^{-1}+\doots &&  &&
&\cr\ilspace
&&{\langle \varphi_2\varphi_3\varphi_4\rangle\langle \varphi_0\varphi_2^2
\rangle\over\langle \varphi_2^2\varphi_4\rangle\langle \varphi_0\varphi_2
\varphi_3\rangle}
&& 1+{1\over2}s^{-1}+\doots && 1+{3\over2}s^{-2}+\doots
&& 1+4s^{-1}+\doots && 0-2s^{-1}+\doots &&  &\cr\ilspace
&&{\langle \varphi_2\varphi_3\varphi_4\rangle\langle \varphi_0\varphi_1^2
\rangle\over\langle \varphi_1^2\varphi_4\rangle\langle \varphi_0\varphi_2
\varphi_3\rangle}
&& 2+27s^{-1}+\doots && 1-{1\over2}s^{-1}+\doots && -2s-33+\doots &&
&& 0+4s^{-2}+\doots &\cr
\tablerule
\noalign{\vskip2pt}
&\multispan{13}\hfil \tenpoint
Table 2: Asymptotic ratios of 3-point functions\hfil\cr
}}$$
}}
\ifx\answ\bigans\box\Bbox\fi

This, in conjunction with the abstract  and general isomorphism we find
between the complex structure moduli space of a Calabi-Yau and the
{\it enlarged}\/
K\"ahler moduli space of its mirror
(using toric geometry), provides us with strong evidence that our
understanding of Calabi-Yau conformal field theory moduli space
is correct. In particular, as our earlier discussion has emphasized,
this implies that the basic operation of deformation by a truly
marginal operator
(from a spacetime
point of view, this corresponds to a slow
variation in the vacuum expectation value of a scalar field with
an exactly flat potential)
can result in a change in the topology of
the Calabi-Yau target space.
 This discontinuous mathematical
change, however, is perfectly smooth from the point of view of physics.
In fact, using mirror symmetry, such an evolution can be reinterpreted
as a smooth, topology preserving,
change in the ``shape'' (complex structure) of the mirror space.

There are two important points we need to mention. First, for ease of
discussion we have focused on the case in which the only
deformations are those associated with the K\"ahler structure
of $M$   and, correspondingly, only the complex structure of
$\tilde M$.
This may have given the incorrect impression that the
topology changing transitions under study can always be
reinterpreted in a topology preserving manner in the mirror
description.
 The generic situation, however, is one in which the complex
structure {\it and}\/ the K\"ahler  structure of $M$ and $\tilde M$
both  change. Again, from a spacetime point of view this simply
corresponds to a slow variation of the expectation values of a set
of scalar fields with flat potentials. Under such circumstances,
topology change can occur in both the original and the mirror
description. Our reasoning will ensure that such changes are
physically smooth. Clearly there is {\it no}\/ interpretation ---
the original or the mirror ---
which can avoid the topology changing character of the processes.

Second, we have used the terms ``enlarged'' and ``partially enlarged''
 in our discussion of the K\"ahler moduli space. We now briefly indicate
the distinction. The central result of the present work
(and that of Witten \rW) is that the proper geometric interpretation of
conformal field theory moduli space requires that we augment
the previously held notion of a single complexified K\"ahler cone
associated with a single topological type of Calabi-Yau space.
In the previous sections we have focused on {\it part}\/ of the
requisite augmentation: we need to include the complexified K\"ahler
cones of Calabi-Yau spaces related to the original by flops
of rational curves (of course, it is arbitrary as to which Calabi-Yau
we call the original). These K\"ahler cones adjoin each other along
common walls. The space so created is the {\it partially}\/ enlarged
K\"ahler moduli space.
 It turns out, though, that conformal field theory
moduli space requires that even more regions be added.  Equivalently,
the partially enlarged K\"ahler moduli space is only a subregion
of the moduli space  which is mirror to the complex structure moduli space
of the mirror Calabi-Yau manifold. The extra regions which need to
be added arise directly from the toric geometric description and
were first identified in
the two dimensional supersymmetric gauge theory approach of
\rW.
 These regions correspond to the moduli spaces of conformal
theories on orbifolds of the original smooth Calabi-Yau,
Landau-Ginzburg orbifolds, gauged Landau-Ginzburg theories and
hybrids of the above.
The union of all of these regions (which also join along common walls)
constitutes the {\it enlarged}\/ K\"ahler moduli space.
 For instance, in the example studied in
section V we found that there were five regions in the partially
enlarged K\"ahler moduli space. The enlarged K\"ahler moduli space,
as it turns out, has $100$ regions. One of these is a Landau-Ginzburg
orbifold region, $27$ of these are sigma models on Calabi-Yau orbifolds,
and $67$ of these are hybrid theories consisting of Landau-Ginzburg
models fibered over various compact spaces.
It is worthwhile emphasizing that in contrast with previously held notions,
orbifold theories are not simply boundary points in the moduli space
of smooth Calabi-Yau sigma models but, rather, they have their own
regions in the enlarged K\"ahler moduli space and hence are more on equal
footing with the smooth examples.

\newsec{Conclusions}

In this talk we have focused on the proper and complete geometric
interpretation of points in the moduli spaces of a class of
$N$=2 superconformal field theories. We have uncovered a surprisingly
rich structure. Previously it was believed that any such moduli space
was interpretable in terms of the complex structure and K\"ahler
structure of an associated Calabi-Yau manifold. We now see that this is
but a small fragment of the full story. The K\"ahler moduli space must
be augmented to the {\it enlarged}\/ K\"ahler moduli space of which
the former is one of many cells adjoined along various common walls.
The new cells correspond to nonlinear sigma models on smooth topologically
distinct Calabi-Yau's (related by flops of rational curves) as well
Calabi-Yau orbifolds, Landau-Ginzburg theories and hybrid models.

There are a number of implications of this augmented picture.
First, we have shown that deformations by truly marginal operators
can take us in a physically smooth manner from any region to any other.
In particular, this means that the topology of the target space
(the universe in a theory of strings) can change with no more exotic
physical impact than mere geometric expansion. Second, the enlargement
of the moduli space harmoniously
clears up some troubling puzzles in mirror symmetry.  More precisely,
whereas it proved difficult to understand how the complexified
K\"ahler moduli space of a Calabi-Yau could be isomorphic to
the complex structure moduli space of its mirror, there is a manifest
isomorphism when the enlarged K\"ahler moduli space is used. Third,
we have seen how we are led to a shift in perspective regarding orbifolds.
Rather than being boundary points in moduli space --- and hence less
than generic --- orbifolds occupy their own regions just like the smooth
Calabi-Yau manifolds. Fourth and finally, we have mentioned that
the enlarged K\"ahler moduli space  generally contains numerous regions
whose most natural interpretation is not in terms of nonlinear sigma
model field theories. The geometric  properties of such models
are presently under study.

\bigbreak\bigskip\bigskip\centerline{{\bf Acknowledgements}}\nobreak
The work of P.S.A.\ was supported by DOE grant
DE-FG02-90ER40542, the work of B.R.G.\ was supported
by the Ambrose Monell Foundation, by a National Young Investigator award and
by the Alfred P. Sloan foundation,
and the work of D.R.M.\ was supported  by NSF grant DMS-9304580
and by an American Mathematical Society Centennial Fellowship.

\listrefs

\bye